\documentclass[fleqn,usenatbib]{mnras}
\usepackage{newtxtext,newtxmath}
\usepackage[utf8]{inputenc}
\usepackage[T1]{fontenc}
\usepackage{ae,aecompl}

\usepackage{graphicx}	
\usepackage{amsmath}	
\usepackage{amssymb}	
\usepackage{float}
\usepackage{bm}

\DeclareFontFamily{U}{wncy}{}
\DeclareFontShape{U}{wncy}{m}{n}{<->wncyr10}{}
\DeclareSymbolFont{mcy}{U}{wncy}{m}{n}
\DeclareMathSymbol{\Sh}{\mathord}{mcy}{"58}

\title[Blind deconvolution in astronomy with AO]{Blind deconvolution in astronomy with adaptive optics:\\ the parametric marginal approach}

\author[R. JL. F{\'e}tick et al.]{
R. JL. F{\'e}tick,$^{1,2}$\thanks{E-mail: romain.fetick@lam.fr}
L. M. Mugnier,$^{2}$
T. Fusco,$^{2,1}$
and B. Neichel$^{1}$
\\
$^{1}$Aix Marseille Univ, CNRS, CNES, LAM, Marseille, France\\
$^{2}$DOTA, ONERA, Universit\'e Paris Saclay, F-92322 Ch\^{a}tillon, France\\
}

\date{Accepted XXX. Received YYY; in original form ZZZ}
\pubyear{2020}

\begin{document}
\label{firstpage}
\pagerange{\pageref{firstpage}--\pageref{lastpage}}
\maketitle


\begin{abstract}
One of the major limitations of adaptive optics (AO) corrected image post-processing is the lack of knowledge on the system point spread function (PSF). The PSF is not always available as a direct imaging on isolated point like objects such as stars. Its prediction using AO telemetry also suffers from serious limitations and requires complex and yet not fully operational algorithms. A very attractive solution consists in a direct PSF estimation using the scientific images themselves thanks to blind or myopic post-processing approaches. We demonstrate that such approaches suffer from severe limitations when a joint restitution of object and PSF parameters is performed. As an alternative we propose here a marginalized PSF identification that overcomes this limitation. Then the PSF is used for image post-processing. Here we focus on deconvolution, a post-processing technique to restore the object, given the image and the PSF. We show that the PSF estimated by marginalisation provides good quality deconvolution. The full process of marginalized PSF estimation and deconvolution constitutes a successful blind deconvolution technique. It is tested on simulated data to measure its performance. It is also tested on experimental adaptive optics images of the asteroid 4-Vesta by VLT/SPHERE/Zimpol to demonstrate application to on-sky data.
\end{abstract}




\section{Introduction}

Ground based astronomy is challenging due to the difficulty to acquire highly resolved images. The major issue comes from the atmospheric turbulence, responsible of inhomogeneities in the propagation medium and resulting in optical aberrations \citep{fried1966,Roddier1981}. Long exposure images are blurred by the turbulence effect and lose their details. The corresponding so-called Point-Spread-Function (PSF) is much larger than the diffraction limit of the biggest current telescopes.\\

Adaptive optics (AO) systems have been developed to mitigate the turbulence by a fast compensation of the atmospheric random fluctuations \citep{rousset1990,Roddier1999}. It results in a sharper PSF and hence in a more resolved image. However the AO correction is partial \citep{conan1994eso,conan1994phd}, especially at short wavelength, and the images still suffer from a residual blurring. Post processing techniques, such as deconvolution \citep{Titterington-85a,demoment1989,idier2008}, can then be performed to restore the blurred information in the image close to its original level. Deconvolution algorithms require the knowledge of the PSF to estimate correctly the object. This is particularly critical in astronomy since the PSF is highly variable due to the random evolution of the turbulence and to the complex dependency of the AO correction quality with respect to the observing conditions \citep{rigaut1998sh,fusco2004stat}. Many observational cases, such as small field of view imaging, do not allow the extraction of the PSF from an isolated bright star in the image. Consequently, other methods must be used to identify the PSF from the image.\\

Blind and myopic deconvolution algorithms try to estimate both the observed object and the PSF. The commonly used approach called joint estimation, consists in a simultaneous estimation of the object and the PSF from a given image. This inverse problem accepts many solutions and must be strongly regularised on both the PSF and the object in order to remove the ambiguity of disentangling them. Regularisation on the object is often performed by a convex penalisation of its gradient. Regarding the PSF, fundamental constraints are its unit sum and positivity, achieved by projection \citep{ayers1988} or reparameterization \citep{thiebaut1995,biraud1969}. In most of the cases these constraints are not sufficient to properly retrieve the PSF. \cite{conan1998} and \cite{mugnier2004} developed MISTRAL, for Myopic Iterative STep-preserving Restoration ALgorithm, where the gradient of the object is regularised with a $l^2-l^1$ norm \citep{rey1983,brette1996,mugnier2001}, and the PSF is regularized by means of a Gaussian prior probability, i.e. with an average guess PSF and a power spectral density (PSD) to control its variations around the average. In practice it works properly when the average guess PSF is close to the true PSF and the PSD is small, however there is no guarantee of convergence \citep{levin2009,Blanco2011}.\\

To ensure consistent estimation, it is critical to reduce the number of degrees of freedom from the object and/or the PSF. The informative content of an observation is proportional to the number of degrees of freedom. A parsimonious parameterization of the object is difficult to provide since we have few prior information on it. However a parsimonious parametric model of the PSF is suited for deconvolution processes \citep{drummond1998,markham1999,fetick2019vesta}. It is thus more efficient to parameterize the PSF using the available information on the observing system. The PSF model developed in \cite{fetick2019} is adapted to AO corrected images since it is able to describe both the coherent core and the turbulent halo of the AO corrected PSF. Even with the strong prior of a parsimonious PSF parameterization, the joint maximum a posteriori estimator is not ensured to converge towards a satisfactory result. Indeed the degeneracy of this estimator was demonstrated by \cite{Blanco2011} with a linear combination of PSFs for biomedical imaging application. We show in this paper that the joint estimator is also degenerated with a non linear PSF model, although adapted to astronomical imaging.\\

Due to the drawbacks of the joint estimator, we consider another method that is the marginal estimator \citep{berger1999,Blanc2003b,levin2009}. This method consists in integrating -- marginalizing -- the object out of the problem to estimate only the PSF parameters. This estimator is asymptotically consistent and non-biased \citep{lehmann1983} when the number of data or the signal-to-noise (SNR) ratio increase. We apply this method on our astronomical case to estimate the PSF parameters. The PSF is then used in a conventional (non-blind and non-myopic) deconvolution to estimate the object.\\

In Section \ref{sec:setup} we motivate our choice of a parametric PSF model adapted to AO corrected images. The first approach would be to use this parametric model to perform joint estimation. In Section \ref{sec:jointestimator} we show that the deconvolution based on a joint estimation of the PSF and the object does not converge towards the correct solution in spite of the parametric PSF model introduced. We thus consider the marginalized approach in Section \ref{sec:marginalestimator}. We study the impact of the signal to noise ratio and the supervision level on the PSF estimation and image deconvolution. We show this method to be adapted for blind deconvolution, with example on astronomical images from the VLT/SPHERE/Zimpol instrument \citep{beuzit2019sphere}. Finally Section \ref{sec:conclusion} concludes our work.

\section{PSF and imaging models}
\label{sec:setup}

\subsection{The PSF model}

Due to the highly variable atmospheric conditions and the complexity of the AO correction, the PSF must be estimated for each observation. The simple method of estimating a PSF on a bright star is not possible for many observational cases, especially for observation of resolved objects in a small field of view. The PSF must be estimated with another method in order to be able to perform post-processing techniques on the image. This estimation might be based on telemetry or on the image itself. Since the complete shape of the PSF is not well known, a parametric model overcomes this issue. The parameterization of the long-exposure PSF models the PSF shape in a parsimonious fashion. The PSF model must satisfy the following qualitative requirements:
\begin{itemize}
    \item Parsimony. A small number of parameters reduces efficiently the number of unknowns to estimate.
    \item Versatility. Be able to adapt to different observing systems and observing conditions.
    \item Accuracy. Describe the PSF with enough precision for the considered application, here the deconvolution.
\end{itemize}
The first requirement is antagonistic with the other ones since a small number of parameters often leads to a rigid model, that might have difficulties to adapt with versatility and accuracy to all the possible PSFs. The use of the physics of the observation into the PSF model partially overcomes this limitation by allowing to distinguish the constant part of the PSF from the free part. In case of astronomical observation, the \cite{Roddier1981} equation states that the PSF is the convolution of the atmospheric PSF and the static PSF, the latter being made of the telescope and the detector contributions
\begin{equation}
    h = h_A \otimes h_T \otimes h_D
\end{equation}
with $\otimes$ the continuous convolution operator, $h$ the total PSF, $h_D$ the detector contribution, $h_T$ the telescope and instrument contribution, and $h_A$ the atmospheric contribution. \cite{conan1994phd} has shown this equation to be still valid for AO corrected turbulence. The telescope PSF is given by the diffraction theory \citep{goodman2005}. It can be derived theoretically from the telescope optical setup (e.g. diameter, focal length), or it can be measured experimentally to take into account possible aberrations \citep{ndiaye2013}. It can also include Non Common Path Aberrations (NCPA) for AO systems, in the case they are known. This instrumental PSF is supposed to be quite well known and static with respect to the highly variable atmospheric PSF. This atmospheric PSF is better described by its Fourier transform, called Optical Transfer Function (OTF). Indeed the turbulent OTF is directly related to the autocorrelation of the phase $\phi$ of the incident electromagnetic wave
\begin{equation}
    \Tilde{h}_A(\Vec{\rho}/\lambda) = e^{-B_\phi(0)}e^{B_\phi(\Vec{\rho})}
\end{equation}
where $\Tilde{h}_A$ denotes the Fourier transform of $h_A$, $\Vec{\rho}$ is the position variable, $\lambda$ is the observation wavelength and $B_\phi$ the autocorrelation of the electromagnetic phase. Moreover the Wiener-Khintchine theorem links the autocorrelation to the Power Spectral Density (PSD) as
\begin{equation}
    W_\phi(\Vec{f}) = \mathcal{F}\left\{ B_\phi(\Vec{\rho}) \right\}
\end{equation}
where $W_\phi$ is the phase PSD, and $\mathcal{F}$ denotes the Fourier operator. The variables $\Vec{f}$ and $\Vec{\rho}$ are Fourier conjugated. Using these equations, we finally write the relation between the phase PSD and the long exposure PSF
\begin{equation}
\label{eq:psd_to_psf}
    h(\Vec{u}) = \mathcal{F}^{-1} \left\{ \Tilde{h}_D(\Vec{\rho}/\lambda) \; \Tilde{h}_T(\Vec{\rho}/\lambda) \; e^{-B_\phi(0)}e^{\mathcal{F}^{-1}\{ W_\phi(\Vec{f}) \}}  \right\}
\end{equation}
where the variables $\Vec{u}$ and $\Vec{\rho}/\lambda$ are Fourier conjugated. The term $B_\phi(0)$ is the integral of $W_\phi$ over the frequency plane. A parameterization of the phase PSD is thus equivalent to a parameterization of the PSF. This relation makes possible to parameterize the PSF from the PSD of the electromagnetic phase, with the substantial advantage to include physical parameters in the PSF model. Indeed the turbulent phase PSD is well described by the \cite{kolmogorov1941} power law in $f^{-11/3}$ and the \cite{fried1966} parameter, called $r_0$, that defines the strength of the turbulence applied to optical observations. We use the AO residual phase PSD model defined in \cite{fetick2019}. This phase PSD model is split into two terms. One term corresponds to the AO corrected core at frequencies seen by the AO system, lower than the AO cutoff frequency $f_\text{AO}$. The second term corresponds to frequencies $f>f_\text{AO}$ not corrected by the AO system and behaving as the complete Kolmogorov turbulence. The first term is the most difficult to parameterize since it includes the complex effect of the AO partial correction. Its shape is based on the \cite{moffat1969} model defined with two parameters $(\alpha,\beta)$ in the axi-symmetric form. The Moffat function writes
\begin{equation}
    M(\Vec{f}) = \left( 1+f^2/\alpha^2 \right) ^{-\beta}
\end{equation}
where $\alpha$ is the elongation in the frequency domain, and $\beta$ is the decrease power law. For a finite energy of the Moffat on the plane, we must ensure $\beta>1$. We consider the following phase PSD model
\begin{equation}
\label{eq:psd_psf_model}
    W_\phi(\Vec{f}) = \left[ \sigma^2 \mathcal{N} M(\Vec{f}) + C \right]_{f\leq f_\text{AO}} + \left[ 0.023 r_0^{-5/3} f^{-11/3} \right]_{f>f_\text{AO}}
\end{equation}
where $r_0$ is the Fried parameter, $\sigma^2$ the phase variance of the Moffat model, $C$ a constant, and $\mathcal{N}$ a normalization factor that ensures a unitary integral of the Moffat over the correction disk $f\leq f_\text{AO}$. It is defined as
\begin{equation}
    \mathcal{N} = \frac{\beta-1}{\pi\alpha^2}\left[ 1-\left( 1+ f^2_\text{AO} / \alpha^2  \right)^{1-\beta} \right]^{-1}
\end{equation}

\subsection{Simulation of an AO corrected PSF}
\label{sec:psf_simulation}

\begin{figure}
\centering
	\includegraphics[width=0.95\columnwidth]{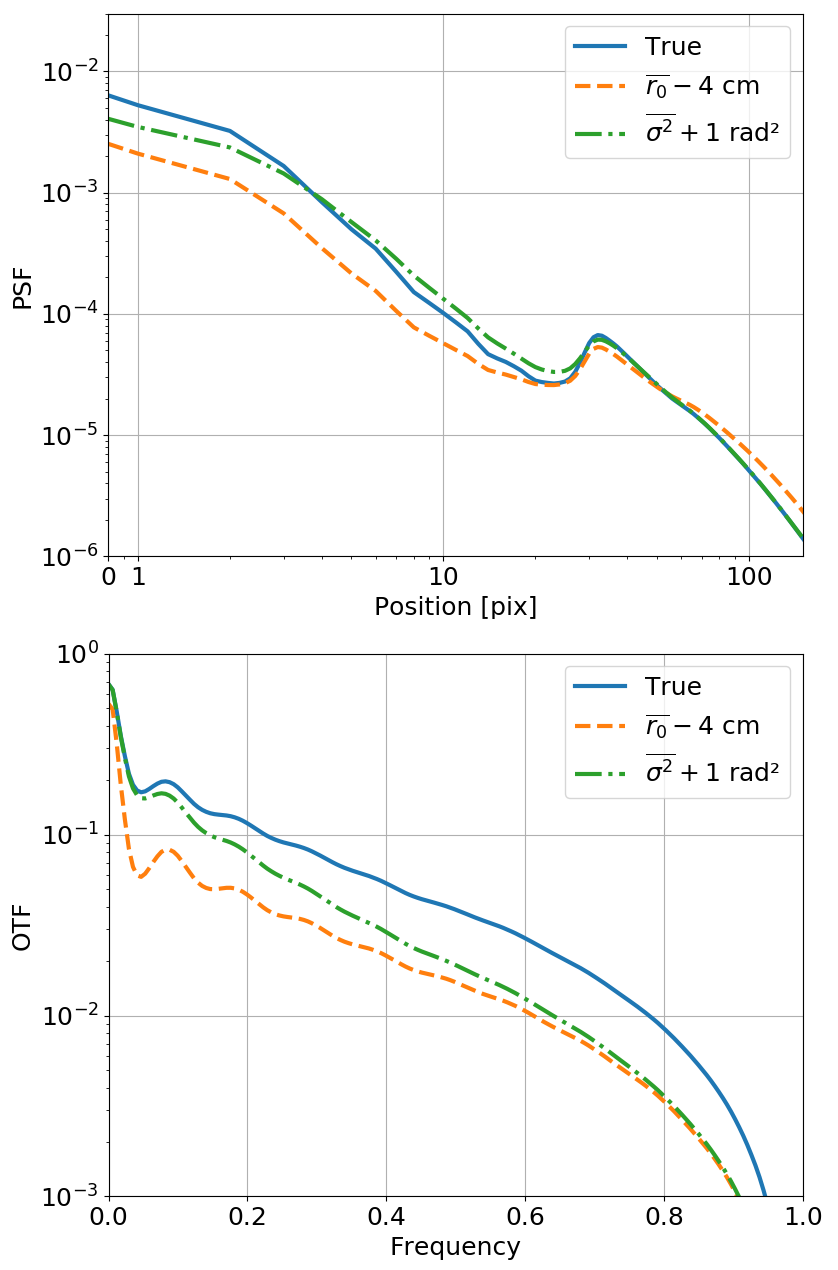}
    \caption{Top: PSF of parameters $(\overline{r_0},\overline{\sigma^2})$ for our simulations (plain line). Impact on the PSF shape of reducing the $r_0$ of an arbitrary value of $4$ cm (dashed). Impact on the PSF shape of increasing the $\sigma^2$ of an arbitrary value of $1$ rad$^2$ (dashed-dotted). Bottom: corresponding OTFs.}
    \label{fig:psf_exemple}
\end{figure}

In the following sections we simulate the observation of a resolved object with an AO system. We use the AO corrected phase PSD model of Eq.~\ref{eq:psd_psf_model} to generate the PSF (Eq.~\ref{eq:psd_to_psf}) corresponding to this observation. The PSF parameters $(\alpha,\beta,C)$ are considered as constant in all the paper since they do not evolve significantly with the observing conditions \citep{fetick2019}. Only the two main parameters $r_0$ scaling the turbulence strength, and $\sigma^2$ scaling the AO residual variance are considered as variable. With this parameterized model, the PSF directly has unit energy on the plane and is positive. These constraints do not have to be implemented for minimization algorithms on the PSF. We only need to ensure the physical constraints $r_0>0$ and $\sigma^2\geq 0$ for $W_\phi$ to be positive and finite. We call $\bm{\gamma}=\{r_0,\sigma^2\}$ the set of parameters to be estimated. The $r_0$ parameter impacts the halo of the PSF an similarly the level of the OTF corrected frequencies $f>0.1$ (Fig.~\ref{fig:psf_exemple}, dashed). The $\sigma^2$ parameter impacts the core of the PSF and the slope of the OTF (Fig.~\ref{fig:psf_exemple}, dashed-dotted).\\

All the parameters and constants used to generate our PSF model are summarised in Table \ref{tab:setup}. We choose $\overline{r_0}=15$ cm and $\overline{\sigma^2}=1.3$ rad$^2$ that corresponds to typical atmospheric condition at Paranal at $\lambda=550$ nm and typical partial AO correction in the visible wavelength. The overline identifies these values as the true PSF values of our simulation. The corresponding Strehl ratio is $\overline{S_R} = 11.5\%$.

\begin{table}
	\centering
	\caption{Setup parameters to define our PSF model. The top section of the table corresponds to the telescope fixed parameters, the middle section corresponds to the turbulent PSF fixed parameters, the bottom section corresponds to the PSF highly variable parameters. The "Lower" column gives the lower bound on each parameter.}
	\label{tab:setup}
	\begin{tabular}{lcccr}
		\hline
		Parameter & Symbol & Value & Lower & Unit\\
		\hline\hline
		Primary diameter & D & 8 & $>0$ & m\\
		Secondary diameter & D' & 1.12 & $\geq 0$ & m\\
		Sampling & $s$ & 3 & $\geq 2$ & N/A\\
		AO cutoff frequency & $f_\text{AO}$ & 1.25 & $>0$ & m$^{-1}$\\
		\hline
		Moffat width & $\alpha$ & 0.05 & $>0$ & m$^{-1}$\\
		Moffat power law & $\beta$ & 1.5 & $>1$ & N/A\\
		AO area constant & $C$ & 0 & $\geq 0$ & rad$^2$m$^2$\\
		\hline
		Fried parameter & $\overline{r_0}$ & 15 & $>0$ & cm\\
		Phase variance & $\overline{\sigma^2}$ & 1.3 & $\geq 0$ & rad$^2$\\
		\hline
	\end{tabular}
\end{table}

\subsection{The imaging equation}

In the isoplanetic domain, the image $i$ is the result of the convolution of the object $o$ with the PSF $h$. The equation of observation on the detector is
\begin{equation}
    \bm{i} = \Sh (h\otimes o) + \bm{n}
\end{equation}
where $\Sh$ is the discretization operator, $\bm{i}$ the discrete image and $\bm{n}$ an additive noise that takes into account photon shot noise and detector read out noise on each pixel. The $\Sh$ function is assumed to take the value of $h\otimes o$ at the centre of each pixel and to multiply it by the pixel area (for flux conservation). We can directly consider the discretized object $\bm{o}$ and the discretized PSF $\bm{h}$ instead of their continuous versions. The discrete equation of observation becomes
\begin{equation}
\label{eq:image}
    \bm{i} = (\bm{h}\star \bm{o})  + \bm{n}
\end{equation}
where $\star$ denotes the discrete convolution operator.

\subsection{The asteroid Vesta as resolved object}

\begin{figure}
\centering
	\includegraphics[width=\columnwidth]{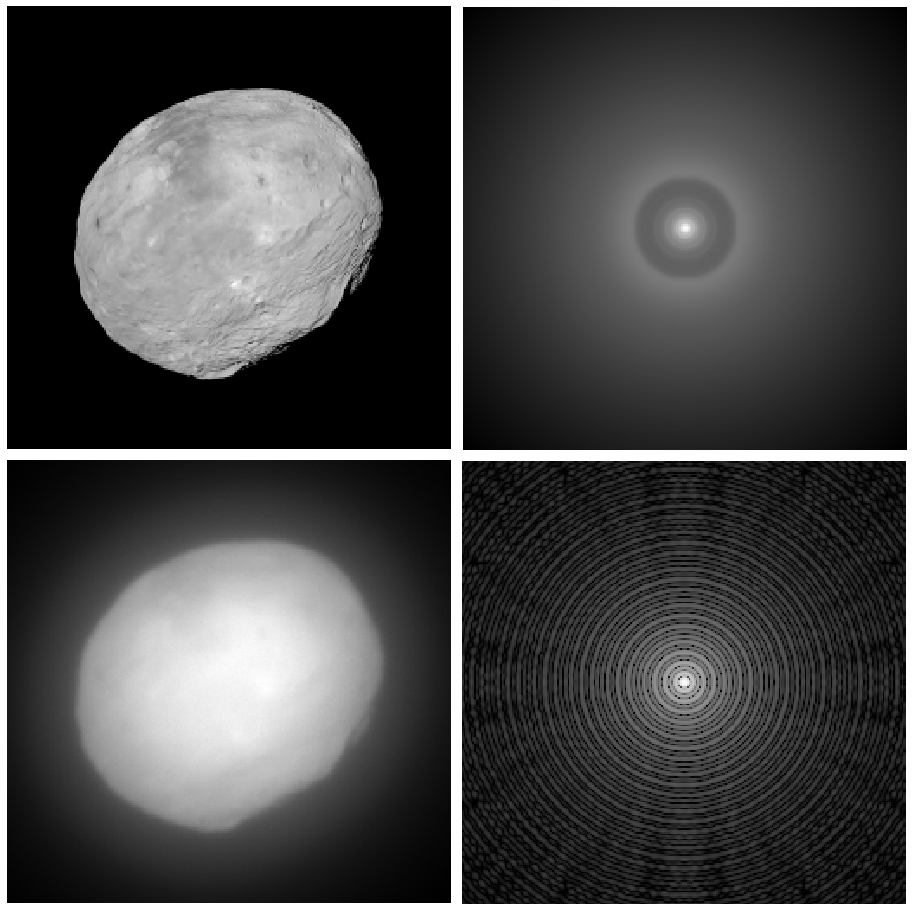}
    \caption{Top left: OASIS model of the asteroid Vesta. Top right: true PSF of the simulation (log intensity scale). Bottom left: observed image. Bottom right: diffraction PSF, called $h_T$ in the text (log intensity scale). Simulation arrays are $512\times 512$ pixels, but are cropped at $256\times 256$ pixels for this figure.}
    \label{fig:simulation}
\end{figure}

We simulate an observation of the Main Belt asteroid Vesta. The asteroid was visited by the NASA/Dawn mission and observed at high resolution \citep{russell2012} with its framing camera. The OASIS \citep{jorda2010} software was then used to build a synthetic view of the asteroid \citep{fetick2019vesta} from the high resolution Dawn data. The resulting synthetic object is shown on Fig.~\ref{fig:simulation} (top left). Equation \ref{eq:image} is then used to derive the image corresponding to the observation of this object with the simulated PSF of parameters $(\overline{r_0},\overline{\sigma^2})$. The Poisson photon noise and the detector read out noise are taken into account. We consider a total flux of $10^{9}$ photons and a detector read out noise of $20$ electron standard deviation. The corresponding signal to noise ratio (SNR) is approximately of $105$. The image is shown on Fig.~\ref{fig:simulation} (bottom left). These values of flux and read out noise are typical from VLT/SPHERE/Zimpol asteroid observations (ESO Large Program ID 199.C-0074), as it will be discussed in Sect.~\ref{sec:sphere}. The simulated images are $512\times 512$ pixels wide.\\

The blurred and noisy image obtained from the simulation of an AO observation has lost the details and sharp edges (high spatial frequencies) of the object. Since the observer is assumed to have no access to the PSF, nor to its $(\overline{r_0},\overline{\sigma^2})$ parameters, a method of estimation of both the PSF parameters and the object must be used to retrieve the details on the object. In Sect.~\ref{sec:jointestimator}, we investigate the conventionally used joint estimation technique, that is an attempt to simultaneously retrieve the PSF and the object. It is different from the marginal method of Sect.~\ref{sec:marginalestimator}, that estimates first the PSF parameters and then the object.

\section{Testing the conventional joint estimation}
\label{sec:jointestimator}


When trying to perform blind deconvolution, the most simple and conventional method is the joint estimation: a simultaneous estimation of both the PSF and the object for a given image. On a probabilistic point of view, one wants to maximise the posterior probability $P(\bm{o},\bm{h}|\bm{i})$ of the object and the PSF, given the image \citep{mugnier2004}. Since the PSF is parameterized with the set $\bm{\gamma}$, the probability is written $P(\bm{o},\bm{\gamma}|\bm{i})$. We use the Bayesian rule to write this posterior probability
\begin{equation}
    P(\bm{o},\bm{\gamma}|\bm{i}) \propto P(\bm{i}|\bm{o},\bm{\gamma}) \; P(\bm{o}) \; P(\bm{\gamma})
\end{equation}
with $P(\bm{i}|\bm{o},\bm{\gamma})$ the likelihood, $P(\bm{o})$ and $P(\bm{\gamma})$ the prior probabilities on the object and the PSF parameters respectively. The conventional solution is the one that maximises $P(\bm{o},\bm{\gamma}|\bm{i})$ and is called Joint Maximum A Posteriori (JMAP). We consider an additive noise, non necessarily stationary over the pixels since it takes into account both read out noise and Poisson shot noise. \cite{mugnier2004} have shown that the total noise on the detector is well approximated by a non stationary Gaussian noise with the contribution of both read-out noise and Poisson noise. The variance of the total noise at pixel $(k,l)$ is
\begin{equation}
\label{eq:noise}
    \left( \bm{\sigma_n^2} \right)_{k,l} = \sigma^2_\text{RON} + \left( \bm{\sigma^2_\text{Poisson}} \right)_{k,l}
\end{equation}
 This noise expression is suited for astronomical cases, over a wide range of SNR. Under this hypothesis, the JMAP problem is equivalent to the minimisation of the expression 
\begin{equation}
    J_\text{JMAP}(\bm{o},\bm{\gamma}) = \left\Vert \frac{\bm{i}-\bm{o}\star \bm{h_\gamma}}{\bm{\sigma_n}} \right\Vert^2 -\ln P(\bm{o}) -\ln P(\bm{\gamma})
\end{equation}
where $J_\text{JMAP}$ is the neg-logarithm of $P(\bm{o},\bm{\gamma}|\bm{i})$, and $\bm{h_\gamma}$ is the PSF model depending on the parameters $\bm{\gamma}$. The prior probability on the object is a regularization term that prevents the well-known dramatic increase of the noise during the deconvolution process, that would lead to unusable results. Different regularizations are possible, such as the $l^2$ (Wiener like) producing smoothed edges, $l^1$ (total variation) producing cartoon-like restored objects. The so-called $l^2-l^1$ regularisation \citep{rey1983,brette1996,mugnier2001} is adapted to restore objects comprising both smooth regions and sharp edges, and has been applied successfully in astronomy. The joint MAP criterion for this $l^2-l^1$ norm \citep{mugnier2004} is defined as
\begin{equation}
\label{eq:map_l1l2}
\begin{split}
    J_\text{JMAP}(\bm{o},\bm{\gamma};\delta,\kappa) = & \left\Vert \frac{\bm{i}-\bm{o}\star \bm{h_\gamma}}{\bm{\sigma_n}} \right\Vert^2 \\
    & + \delta^2\sum_{k,l} \left\vert \frac{\nabla \bm{o}_{k,l}}{\delta\kappa} \right\vert -\ln\left( 1+\left\vert \frac{\nabla \bm{o}_{k,l}}{\delta\kappa} \right\vert \right)\\
    & - \ln{P(\bm{\gamma})}
\end{split}
\end{equation}
where $\kappa$ and $\delta$ are the two regularization hyper-parameters which must be set before the minimisation, and $\nabla$ is the isotropic gradient operator defined as
\begin{equation}
    \nabla \bm{o}_{k,l} = \sqrt{(\bm{o}_{k,l}-\bm{o}_{k,l-1})^2 + (\bm{o}_{k,l}-\bm{o}_{k-1,l})^2}
\end{equation}
The $\kappa$ parameter is a scaling factor of the object's gradient that acts as the inverse of a global hyper-parameter, and that has the same unit as the object (photo-electrons). The $\delta$ parameter is a unitless threshold between the quadratic behaviour that filters low amplitude noise and the linear behaviour that allows high amplitude sharp edges. For $\delta\to 0$ the regularisation tends towards the total variation norm, for $\delta\to\infty$ the regularisation tends towards a $l^2$ norm. These two hyper-parameters must be provided in a supervised fashion by the user. We use the empirical values
\begin{equation}
    \left\{
    \begin{array}{rl}
     \kappa & =  5 \sqrt{\frac{1}{N}\sum_{k,l}(\bm{i}_{k,l}-\bm{i}_{k,l-1})^2 + (\bm{i}_{k,l}-\bm{i}_{k-1,l})^2} \\ 
     \delta & = 0.4
    \end{array}
    \right.
\end{equation}
with $N$ the number of pixels in the image. These values are chosen to provide a correct deconvolution when the deconvolving PSF is the true PSF. The same values of $(\kappa,\delta)$ are used for all our simulations.\\ 

\begin{figure}
\centering
	\includegraphics[width=.9\columnwidth]{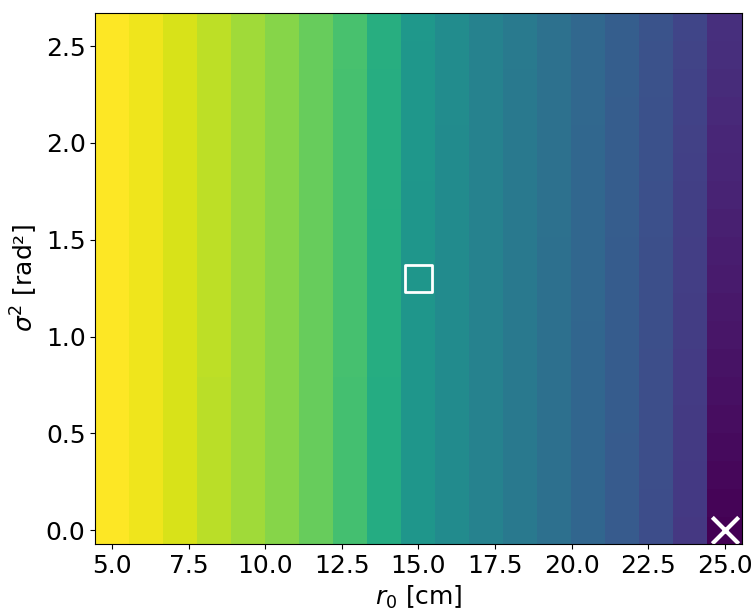}
    \caption{Joint criterion $J_\text{JMAP}$ (log intensity scale) as function of the $r_0$ and $\sigma^2$ PSF parameters. Central square indicates the true PSF parameters, cross indicates the minimum of the criterion.}
    \label{fig:Jmap}
\end{figure}

\begin{figure}
\centering
	\includegraphics[width=\columnwidth]{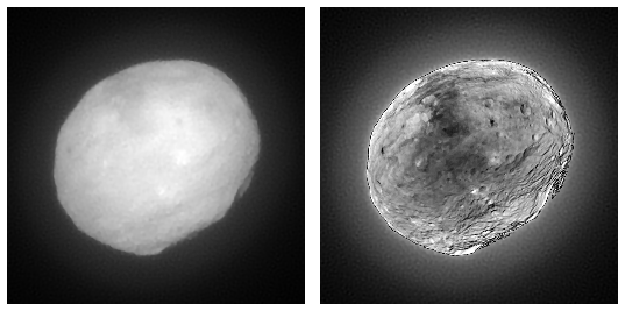}
    \caption{Left: deconvolution of the image using the PSF estimated by the joint estimation. The deconvolved object shows little detail. Right: two times the absolute difference with the true object.}
    \label{fig:deconv_jmap}
\end{figure}

There is no analytical solution to find the minimum of the criterion, and a fast quasi-Newton iterative descent algorithm \citep{thiebaut2002} is used in our case to solve this problem. In addition, constraint of positivity on the object is implemented by projection at each iteration of the algorithm. The prior probability on the PSF parameters is considered as uniform, that is the least informative of all the density probabilities. The uniform probability is simply implemented as upper and lower bounds in the minimizer. The range for the Fried parameter $r_0$ is $5$ cm to $25$ cm, the range for the AO variance $\sigma^2$ is $0$ rad$^2$ to $2.6$ rad$^2$. The value of the criterion $J_\text{JMAP}$ at convergence is shown on Fig.~\ref{fig:Jmap} for different values of PSF parameters $r_0$ and $\sigma^2$. It exhibits a strong evolution with $r_0$ and a soft evolution with $\sigma^2$. The global minimum of the criterion is located at the maximal $r_0=25$ cm and the minimum $\sigma^2=0$ rad$^2$, this is the sharpest PSF possible in the given range. The joint estimator is thus degenerated: it is very far, and actually independent, of the true value of the PSF parameters. It confirms and follows the previous results of \cite{Blanco2011} that found the joint estimator to be degenerated in case of a $l^2$ Wiener-like regularization and no positivity on the object. Our attempt to use the positivity, the $l^2-l^1$ norm adapted to sharp edged objects and the PSF model adapted to AO is not able to remove this degeneracy. An explanation of the joint estimation degeneracy was formulated in \cite{levin2009}: the data fidelity term as $\bm{o}\star \bm{h_\gamma}$ accepts indifferently multiple $(\bm{o},\bm{h_\gamma})$ solutions, while the regularisation favours smooth objects. The result is an over-smoothed estimated object (Fig.~\ref{fig:deconv_jmap}) and a over-sharp estimated PSF. The error on the deconvolved object is defined as the $l^1$ difference between the deconvolved object and the true object, normalised by the flux
\begin{equation}
\label{eq:obj_error}
    \varepsilon_\text{obj} = \frac{\left\Vert \bm{\hat{o}} - \bm{o} \right\Vert_1}{\sum_{k,l} o_{k,l}} = \frac{{\sum_{k,l}\left\vert\hat{o}_{k,l} - o_{k,l}\right\vert}}{\sum_{k,l} o_{k,l}}
\end{equation}
where $\hat{o}$ is the deconvolved object and $o$ the true object. For the joint MAP estimation, we find $\varepsilon_\text{obj,JMAP} = 52\times 10^{-4}$. That is a high error, since the deconvolution of the image by the true PSF gives $\varepsilon_\text{obj,true PSF} = 3.4\times 10^{-4}$. There is approximately a factor of $\sim 15$ in the loss of quality for the given $\varepsilon$ norm. The JMAP deconvolution consequently appears to be of poor quality.\\

\section{The marginal approach}
\label{sec:marginalestimator}

\subsection{Expression of the marginal estimator}

The failure of the joint estimation comes from the difficulty to disentangle the object from the PSF in the image. The high number of unknowns and the coupling of the PSF and the object in the image is responsible of the degeneracy. The number of unknowns should be drastically reduced to hopefully be able to perform correctly blind deconvolution. The marginal approach \citep{Blanc2003b} consists in integrating some variables over their probability. For identification of the PSF parameters we marginalize over the object
\begin{equation}
    P(\bm{\gamma}|\bm{i}) \propto P(\bm{i}|\bm{\gamma}) \; P(\bm{\gamma}) = \int_\Lambda P(\bm{i}|\bm{o},\bm{\gamma}) \; P(\bm{\gamma}) \; P(\bm{o}) \; \text{d}\bm{o}
\end{equation}
where $\Lambda$ is the space of all the possible realisations of the object. This drastically reduces the number of unknowns to estimate (typically a quarter of million for $512\times 512$ images) to a few PSF parameters. Moreover as the number of pixels increases, the number of PSF parameters remains constant, thus increasing the statistical contrast: ratio of the number of measurements over unknowns. This is not the case for the joint estimation, since the statistical contrast remains nearly constant and greater than $1$ when the number of pixels increases. In order to analytically compute the marginalisation integral we make the assumption that the object has a Gaussian probability law of average $\bm{o_m}$ and covariance matrix $R_o$. Since the noise (Eq.~\ref{eq:noise}) and the object and are assumed to be Gaussian, it leads to an image with Gaussian statistics too. Its probability is written
\begin{equation}
    P(\bm{i}|\bm{\gamma}) \propto \frac{1}{\sqrt{\det R_i}}e^{-(\bm{i}-\bm{i_m})^tR_i^{-1}(\bm{i}-\bm{i_m})/2}
\end{equation}
where $\bm{i_m}=\bm{h_\gamma}\star \bm{o_m}$ for a zero mean noise, and $R_i$ is the covariance matrix of the image, depending on the covariance of the object and the noise. Although typical images are combinations of photon, read out noise and potentially background noise, we make the simplification hypothesis of a stationary noise \citep{Blanco2011} in order to be able to write a simple criterion in the Fourier domain for a numerical minimisation in a reasonable amount of time. With this hypothesis of a Gaussian stationary noise, the image covariance matrix writes
\begin{equation}
    R_i=H_\gamma R_oH_\gamma^t + \langle\sigma_n^2\rangle \, I_d
\end{equation}
where $\langle\sigma_n^2\rangle$ is the average noise variance over the pixels, $I_d$ is the identity matrix, and $H_\gamma$ is the operator associated to the discrete convolution by the PSF $\bm{h_\gamma}$. Taking the neg-logarithm of $P(\bm{\gamma}|\bm{i})$ we write the marginal criterion in the Fourier domain \citep{Blanco2011}
\begin{equation}
\label{eq:marginal}
\begin{split}
    J_\text{marg}(\bm{\gamma};S_\text{obj},\langle\sigma_n^2\rangle) = & \sum_f\ln\left( S_\text{obj}(f) \, \vert\Tilde{\bm{h_\gamma}}(f)\vert^2 + \langle\sigma_n^2\rangle \right) \\
     & + \sum_f \frac{\vert \Tilde{\bm{i}}(f)-\Tilde{\bm{h_\gamma}}(f) \, \Tilde{\bm{o}}_m(f)\vert^2}{S_\text{obj}(f) \, \vert\Tilde{\bm{h_\gamma}}(f)\vert^2 + \langle\sigma_n^2\rangle}\\
     & - \ln P(\bm{\gamma})
\end{split}
\end{equation}
with $S_\text{obj}$ the power spectral density of the object. Any prior information on the PSF parameters can be provided through the $P(\bm{\gamma})$ term. Similarly to the assumption we made for the joint estimator, we consider a uniform prior probability over the ranges $5\leq r_0\leq 25$ cm and $0\leq \sigma^2 \leq 2.6$ rad$^2$. This is again the least informative prior for the PSF on these given ranges. Regarding the average object $\bm{o_m}$, we define it as a constant object with the same flux as the image. It is thus the least informative object (the least structured) for the given flux.\\

\begin{figure}
\centering
	\includegraphics[width=\columnwidth]{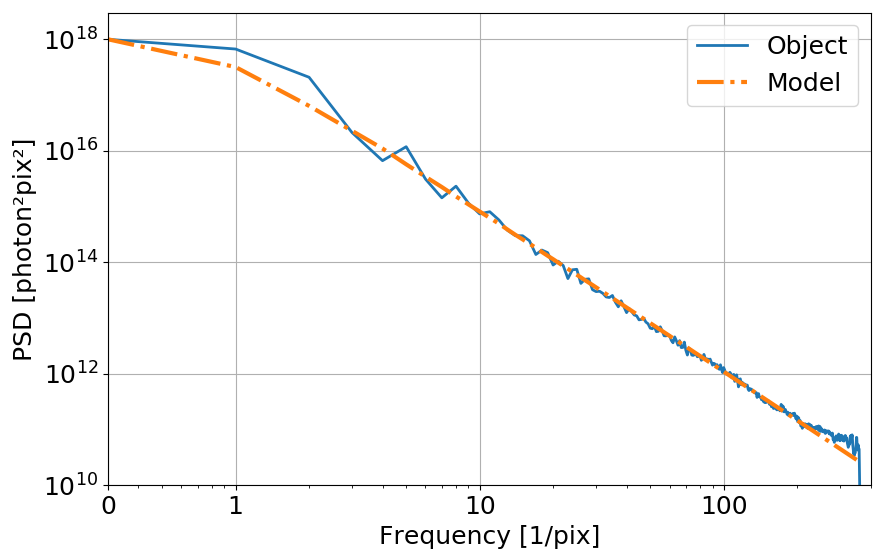}
    \caption{Circular average of the squared modulus of the Vesta Fourier transform (plain blue) and the object PSD model (dotted-dashed orange) for the supervised values $(\overline{k},\overline{\rho_0},\overline{p})$.}
    \label{fig:psdmodel}
\end{figure}

The marginal estimator $J_\text{marg}(\bm{\gamma};S_\text{obj},\langle\sigma_n^2\rangle)$ requires the knowledge of the object PSD and the noise statistics $\langle\sigma_n^2\rangle$. We approximate the PSD of the object with a model \citep{conan1998} depending on three parameters $k$, $\rho_0$ and $p$ in order to reduce the number of unknown to estimate. This object PSD model is an axi-symmetric function in the frequency domain
\begin{equation}
\label{eq:psdmodel}
    S_\text{obj}(f) = \frac{k}{1+(f/\rho_0)^p}
\end{equation}
where $k$ is the object PSD value at $f=0$ (in practice, close to the square of the flux), $\rho_0$ is inversely proportional to the characteristic size of the object, and $p$ is the decrease power law. For example, asteroids have a power law $p\simeq 3$. Artificial satellites have much sharper edges, and a power law $p\simeq 2.5$. An homogeneous field of stars has a power law $p\simeq 0$. The power law is thus an indicator of the sharpness of an object, it consequently defines its category. Figure \ref{fig:psdmodel} shows the circular average of the squared modulus of the asteroid Fourier transform (called "empirical PSD" hereafter, plain blue line on the figure). It also shows the PSD model (dotted-dashed orange line) for $\overline{k}=10^{18}$ square photons, $\overline{\rho_0}=0.9$ pix$^{-1}$ and $\overline{p}=2.91$, where the overline identifies them as the supervised values of the object's PSD model. The supervised choice of these values fits well the object's empirical PSD and will be discussed in Sect.~\ref{sec:sensitivity}. The value of $k$ is only related to the flux of the object and gives no information on its shape. \cite{Blanco2011} define $\mu=\langle\sigma_n^2\rangle/k$ and have found an analytical solution for $k$ given the other parameters. In practice, the criterion to minimise reduces to $J_\text{marg}(\bm{\gamma};\mu,\rho_0,p)$. The PSF model and its gradient with respect to the $\bm{\gamma}$ parameters are provided to the algorithm. The gradients with respect to the hyper-parameters are also provided. The criterion is minimized with the Variable Metric method Limited Memory with Bounds constraints (VMLM-B) \citep{thiebaut2002} descent algorithm. Thanks to this fast algorithm, typical computation time is less than one minute for images of $512\times 512$ pixels on a personal computer. The bounds ensure the strict positivity of the hyper parameters, even though we never reached these bounds. Upper and lower bounds are also provided for the PSF parameters, they are the actual implementation of the uniform prior for this parameters.\\

\subsection{The supervised marginal estimation}
\label{sec:marginal_supervised}

\begin{figure}
\centering
	\includegraphics[width=.95\columnwidth]{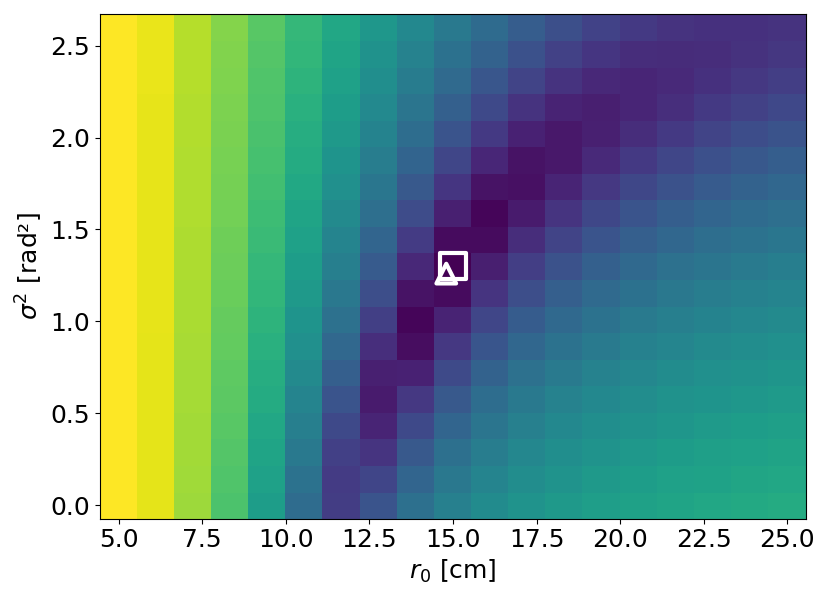}
    \caption{supervised marginal criterion $J_\text{marg}$ (log intensity scale) as function of the $r_0$ and $\sigma^2$ PSF parameters. The parameters $(\mu,\rho_0,p)$ are all fixed to their supervised values $(\overline{\mu},\overline{\rho_0},\overline{p})$. Central square indicates the true PSF parameters, triangle indicates the minimum of the criterion.}
    \label{fig:Jmarg_s}
\end{figure}

In this subsection we present first results of PSF estimation and image restoration using the marginal estimator. This estimation is \emph{supervised}, in the sense that the hyper-parameters $(\mu,\rho_0,p)$ are fixed by the user, similarly to what is done with the joint estimator for parameters $\kappa$ and $\delta$ in Eq.~\ref{eq:map_l1l2}. The result of the estimator depends on the value of these hyper parameters $(\mu,\rho_0,p)$, on the SNR and on the number of data (pixels). The fundamental property of this supervised estimator is to be asymptotically consistent and unbiased \citep{lehmann1983} for increasing SNR or number of data. Here the hyper parameters are fixed to $(\overline{\mu},\overline{\rho_0},\overline{p})$. For our simulated image of the asteroid Vesta with a flux of $10^9$ photons (Fig.~\ref{fig:simulation}, bottom left), the marginal algorithm converges towards the following PSF parameters
\begin{equation}
    \left\{
    \begin{array}{rll}
     r_0 & = 14.8 & \text{ cm} \\ 
     \sigma^2 & = 1.26 & \text{ rad}^2 \\
    \end{array}
    \right.
\end{equation}
The error on the Fried parameter is only $\Delta r_0 = 0.2$ cm, corresponding to a relative error $\Delta r_0/\overline{r_0}=1.3\%$. The error on the AO variance is only $\Delta\sigma^2 = 0.04$ rad$^2$, corresponding to a relative error $\Delta\sigma^2/\overline{\sigma^2}=3.1\%$. The errors on the parameters are small, the residual error of a few percents can originate from
\begin{itemize}
    \item Presence of noise in the image;
    \item Hypothesis of an object with Gaussian statistics;
    \item Over-simplified object PSD model, made of three parameters only;
    \item The supervised PSD parameters are not exactly the optimal ones for the marginal estimator (discussed in Sect.~\ref{sec:sensitivity});
    \item Hypothesis of a stationary noise over the pixels, which is a simplifying assumption.
\end{itemize}
These are thought to be the main sources of residual error for our marginal estimation. Nevertheless, it gives highly satisfactory results. The value of the marginal criterion for different $r_0$ and $\sigma^2$ is seen on Fig.~\ref{fig:Jmarg_s}. The minimum is correctly located near the true values of the PSF parameters. Moreover, the shape on the marginal criterion evolves roughly as the shape of the Strehl ratio (Fig.~\ref{fig:strehl}, triangle denoted \emph{s} for supervised) with the PSF parameters, meaning that a small error on the marginal criterion leads to a PSF with similar Strehl ratio as the true PSF. The Strehl ratio corresponding to the supervised marginal estimation is $S_R=11.4\%$, whereas the Strehl ratio of the true PSF is $\overline{S_R}=11.5\%$. The error on the Strehl ratio is small. Moreover, the shapes of both estimated PSF and OTF match well the true ones (Fig.~\ref{fig:psf_otf}).\\

\begin{figure}
\centering
	\includegraphics[width=\columnwidth]{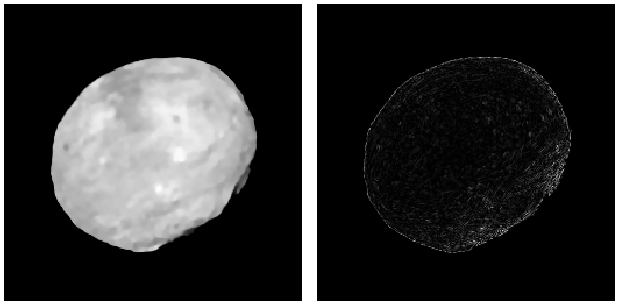}
    \caption{Left: deconvolution of the image using the PSF estimated by the supervised marginal estimation. Right: two times the absolute difference with the true object.}
    \label{fig:deconv_marg_s}
\end{figure}

The deconvolution using a $l^2-l^1$ regularisation \citep{mugnier2004} on the object, is then run in classical (non-myopic) mode with the PSF estimated by marginalisation. It corresponds to Eq.~\ref{eq:map_l1l2} for fixed $\bm{\gamma}$ parameters, since they are provided by the marginal estimation (and not estimated jointly). The result of the deconvolution is shown on Fig.~\ref{fig:deconv_marg_s}. The object is accurately restored, up to the regularisation smoothing effect. The edges are sharp, the details of albedo and craterization are visible. The errors are located mainly on the edges of the object, where are the highest gradients. There is no residual halo, all the energy is restored inside the bounds of the object. No halo and sharp edges are important for precise contour extraction, and thus for a precise volume estimation by astronomers. The error on the deconvolution (Eq.~\ref{eq:obj_error}) with the supervised marginal PSF is $\varepsilon_\text{obj,marg s} = 3.7\times 10^{-4}$. It is close to the error $\varepsilon_\text{obj, true PSF} = 3.4\times 10^{-4}$ of the deconvolution with the true PSF (computed in Sect.~\ref{sec:jointestimator}). The deconvolution with the supervised marginal PSF is nearly identical to the ultimate possible deconvolution. \\

Whereas the joint estimator has shown degeneracy, the supervised marginal estimator exhibits in practice the consistency that is expected from the theoretical properties of this estimator. The error on the estimated PSF is small, and the resulting deconvolution is of high quality.

\subsection{Effect of the SNR on the marginal estimation}

\begin{figure}
\centering
	\includegraphics[width=\columnwidth]{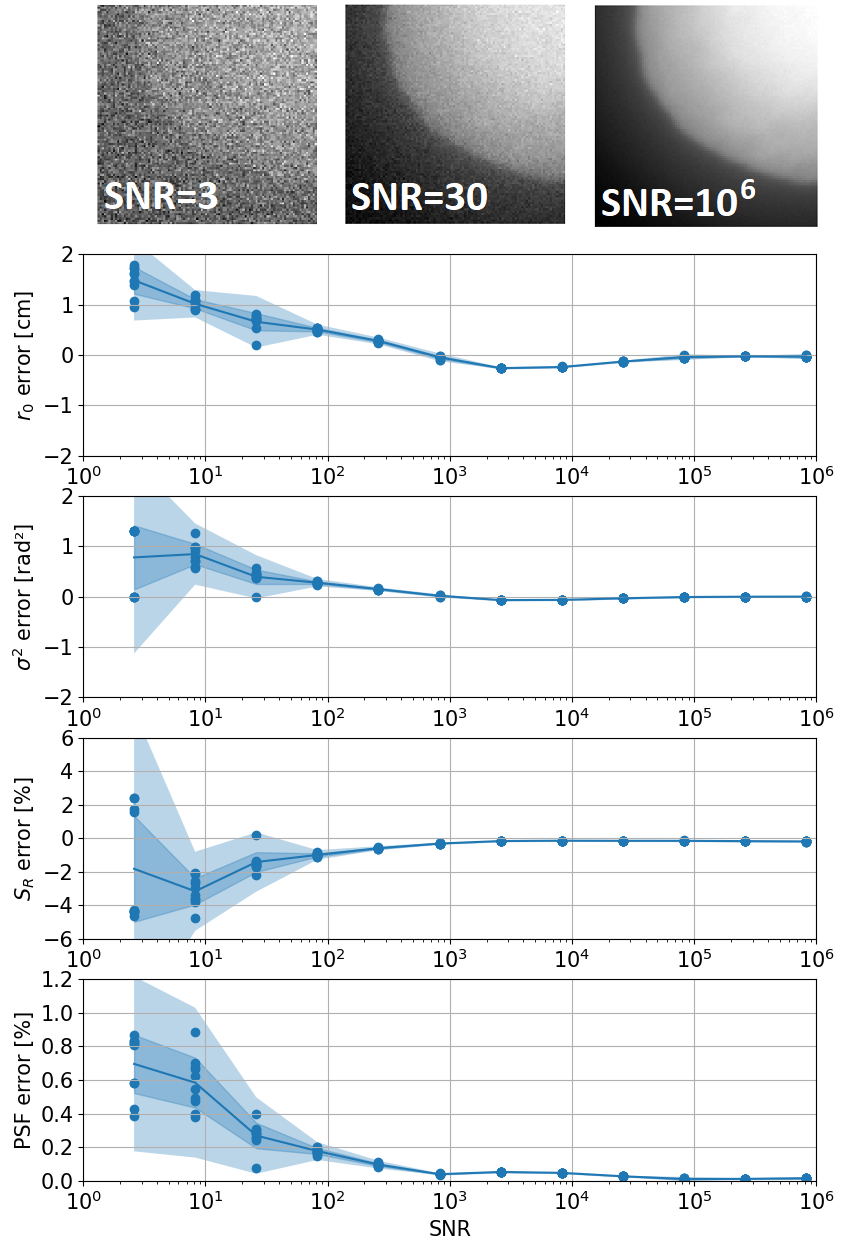}
    \caption{Error on the estimated PSF as function of the signal to noise (SNR) ratio. Top: error on the Fried parameter $r_0$. Second: error on the AO variance $\sigma^2$. Third: Error on the Strehl ratio $S_R$. Bottom: $\varepsilon_h$ error (Eq.~\ref{eq:psferror}) on the PSF shape. For each SNR level, the marginal algorithm has been run 10 times corresponding to 10 samples of noise (dots). The line is the average of the samples per SNR. Shaded areas correspond to the standard deviation (dark) and three times the standard deviation (light). All the hyper-parameters are fixed to their supervised values $(\overline{\mu},\overline{\rho_0},\overline{p})$. At the very top part of the figure, the same detail of the image is shown for three different SNR levels.}
    \label{fig:snr}
\end{figure}

The signal to noise (SNR) ratio is of high importance for inverse problems, including PSF identification and deconvolution. We study the impact of the SNR in our simulation by an increase or a decrease of the a simulated stationary Gaussian noise. We thus define 12 SNR levels, from SNR$\simeq 2.5$ to SNR$\simeq 10^6$. For each SNR case we generate 10 different noise samples. For each of these $12\times 10$ simulated observations, we run the supervised marginal algorithm in order to retrieve the estimated $r_0$ and $\sigma^2$ parameters. The parameters are compared with respect to their true values $\Delta r_0=r_0-\overline{r_0}$ and $\Delta\sigma^2 = \sigma^2 - \overline{\sigma^2}$. The results are shown on Fig.~\ref{fig:snr}. For SNR less than $\sim 30$, there is a strong dispersion due to the noise. The algorithm falls in minima (local or global) generated by the noise. Moreover there is a tendency at low SNR to over-estimate both $r_0$ and $\sigma^2$, and to under-estimate of the Strehl ratio. For SNR higher than $\sim 30$, the noise has little impact on the marginal estimator. For SNR greater than $\sim 300$ the dispersion due to the noise is not visible any more. The marginal estimator converges towards a nearly null errors on the $r_0$ and $\sigma^2$ at high SNR. It results in a nearly null error for the estimation of the Strehl ratio. We also define the $l^2$ error between the estimated PSF shape and the true PSF shape
\begin{equation}
\label{eq:psferror}
\begin{split}
    \varepsilon_h & = \left\Vert \bm{h_\gamma}(r_0,\sigma^2) - \bm{h_\gamma}(\overline{r_0},\overline{\sigma^2})\right\Vert \\ 
    & = \sqrt{\sum_{k,l}\left( \bm{h_\gamma}(r_0,\sigma^2)_{k,l} - \bm{h_\gamma}(\overline{r_0},\overline{\sigma^2})_{k,l} \right)^2}
\end{split}
\end{equation}
where sum runs over the pixels $(k,l)$. This error on the PSF shape is also nearly null for high SNR. Since the $l^2$ norm is conserved by Fourier transform, the error on the OTF shape identically becomes nearly null at high SNR.

\subsection{Sensitivity to the hyper parameters}
\label{sec:sensitivity}

\begin{figure}
\centering
	\includegraphics[width=\columnwidth]{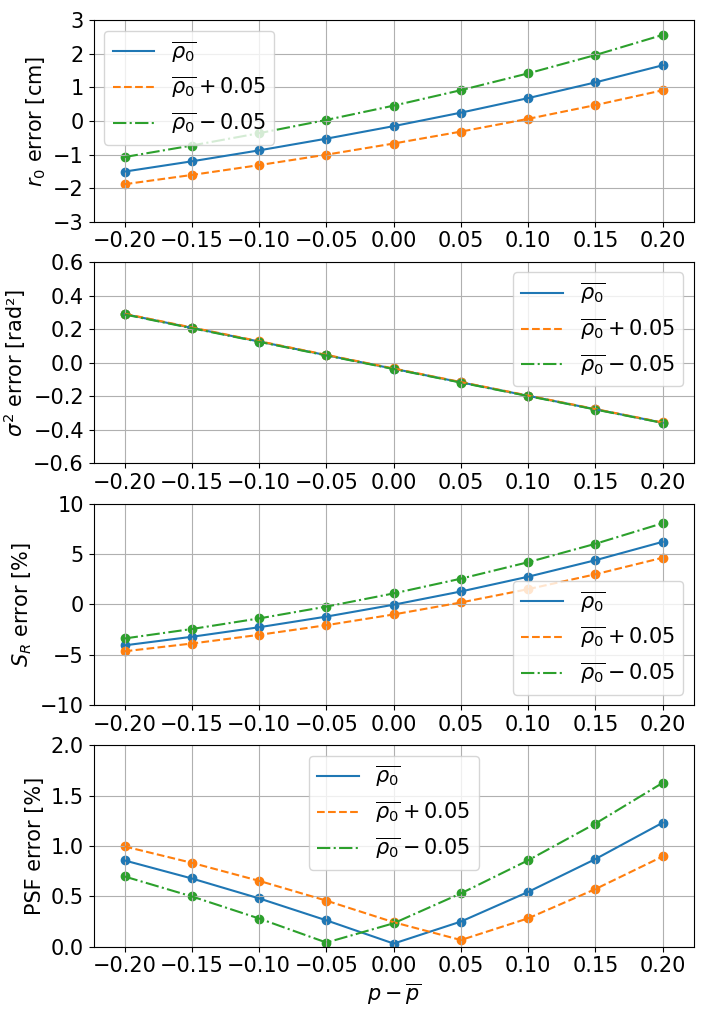}
    \caption{Error on the estimated PSF parameters $r_0$ (top), $\sigma^2$ (second), on the Strehl ratio (third) and the $\varepsilon_h$ error (Eq.~\ref{eq:psferror}, bottom) as function of $p-\overline{p}$. The hyper-parameter $\rho_0$ is set at its fitted value $\overline{\rho_0}$ (plain line), $\rho_0$ is over-estimated (dashed) or under-estimated (dashed-dotted) of $0.05$ with respect to $\overline{\rho_0}$. For all cases, the hyper-parameter $\mu$ is fixed to $\overline{\mu}$.}
    \label{fig:sensitivity}
\end{figure}

The marginal criterion is dependent of the hyper parameters chosen by the user since they affect the outcome of the algorithm. While optimally chosen hyper parameters makes the algorithm to converge to the correct solution, poorly chosen hyper-parameters may bias the result. Figure \ref{fig:sensitivity} shows the sensitivity -- evolution of the error -- of the estimated PSF parameters with respect to the variations of the object PSD power $p-\overline{p}$. The $r_0$ (resp. $\sigma^2$) estimation suffers from a drift of approximately $1$ cm (resp. $-0.2$ rad$^2$) per $0.1$ variation on the object PSD power $p$. A variation of $0.05$ on the $\rho_0$ parameter induces approximately an error of $0.6$ cm on the $r_0$, but has nearly no effect on the $\sigma^2$. Indeed the $\sigma^2$ governs mainly the slope of the OTF (Sect.~\ref{sec:psf_simulation} and Fig.~\ref{fig:psf_exemple}), whereas the $p$ governs the power (slope in log-log plot) of the decrease of the object's spectrum. The parameter $p$ is thus an indicator of the sharpness of the object. The $r_0$ and $\sigma^2$ parameters are adjusted by the marginal criterion to match the shape of the image's spectrum: an increase (resp. decrease) in the object power results in a decrease (resp. increase) of the estimated OTF slope. Regarding the $r_0$ parameter, it is responsible of a general downward offset of the OTF with respect to the null frequency. The value of the object PSD at high frequency evolves as $S_\text{obj}(f)\simeq k(\rho_0/f)^p$. For a given $k$, both $\rho_0$ and $p$ impact the value of the high frequencies, and thus the estimation of the $r_0$ parameter.\\

At $p=\overline{p}$ and $\rho_0=\overline{\rho_0}$, the error on the estimated PSF parameters is nearly null. The residuals are the ones discussed in Sect.~\ref{sec:marginal_supervised}, supposed to originate mainly from to the non Gaussian statistics of the object. The parameters we have chosen for the supervised marginal approach are close to the optimal ones. Nevertheless the residual error on $r_0$ and $\sigma^2$ at $p=\overline{p}$ and $\rho_0=\overline{\rho_0}$ is negligible for our applications, and leads to satisfactory estimation of the PSF shape and to a correct deconvolution.\\

A non optimal choice of hyper-parameters leads to a biased estimation of the PSF parameters. For our applications shifts of $\pm 0.05$ on both $p$ and $\rho_0$ lead to reasonable errors on the PSF parameters of $\pm 0.1$ rad$^2$ on the $\sigma^2$ and approximately half a centimetre on the $r_0$. Instead of providing these hyper parameters in a supervised fashion, it would be useful for practical applications to estimate them simultaneously with the PSF parameters, this constitutes the fully unsupervised method presented in the following subsection.

\subsection{The fully unsupervised marginal estimation}

\begin{figure}
\centering
	\includegraphics[width=.95\columnwidth]{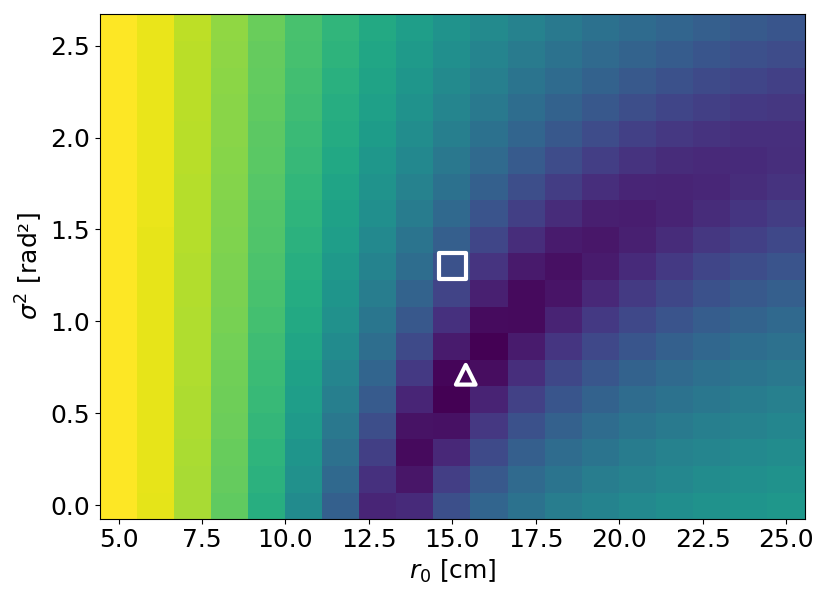}
    \caption{Fully unsupervised marginal criterion $J_\text{marg}$ (log intensity scale) as function of the $r_0$ and $\sigma^2$ PSF parameters. The parameters $(\mu,\rho_0,p)$ are the ones estimated from marginal minimization. Central square indicates the true PSF parameters, triangle indicates the minimum of the criterion.}
    \label{fig:Jmarg_u}
\end{figure}

One can consider estimating the hyper-parameters $(\mu,\rho_0,p)$ along with the PSF parameters in the marginal estimator. This flavor will be called the fully unsupervised marginal estimator and is studied in this Subsection. It means a complete ignorance of the observed object, it is only supposed to be described roughly by the object PSD model of Eq.~\ref{eq:psdmodel}. This model encompasses object as dissimilar as star fields, asteroids, planets, or artificial satellites. For our simulation on the asteroid Vesta, we run the marginal criterion of Eq.~\ref{eq:marginal} to find the PSF parameters and the object PSD parameters, it corresponds to a fully unsupervised marginal estimation. The algorithm converges towards
\begin{equation}
    \left\{
    \begin{array}{rll}
     r_0 & = 15.4 & \text{ cm} \\ 
     \sigma^2 & = 0.71 & \text{ rad}^2 \\
     \mu & = 2.9\times 10^{-9} & \\
     \rho_0 & = 1.49 & \text{ pix}^{-1} \\
     p & = 3.27 &
    \end{array}
    \right.
\end{equation}
The corresponding map of the criterion is given Fig.~\ref{fig:Jmarg_u}. The $r_0$ parameter is over-estimated while the $\sigma^2$ parameter is under-estimated, the estimated PSF has higher Strehl ratio than the true one (Fig.~\ref{fig:strehl}) and a sharper peak (Fig.~\ref{fig:psf_otf}). It results in $p=3.27$ instead of $\overline{p} = 2.91$, i.e. an object's PSD with less energy than the true one at high spatial frequencies.\\

\begin{figure}
\centering
	\includegraphics[width=\columnwidth]{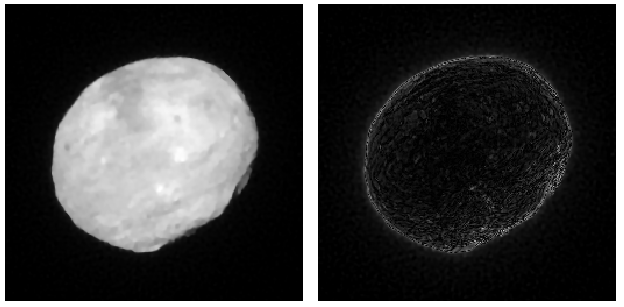}
    \caption{Left: deconvolution of the image using the PSF estimated by the fully unsupervised marginal estimation. Right: two times the absolute difference with the true object.}
    \label{fig:deconv_marginal_all_free}
\end{figure}

The object deconvolved with the fully unsupervised marginal PSF is shown on Fig.~\ref{fig:deconv_marginal_all_free} and is still similar to the object restored in the supervised marginal case: details of albedo and craterization are visible. Edges are sharp, only remains a small residual halo. Since the $r_0$ is slightly over estimated, the PSF halo is slightly under-estimated. The difference between the true halo and the estimated one remains in the deconvolved object. The error on the deconvolution with the fully unsupervised marginal PSF is $\varepsilon_\text{obj,marg u} = 8.1\times 10^{-4}$.\\

Even though the fully unsupervised marginal estimator is more stable than the joint estimator, the resulting object is still over-smoothed, due to the difficulty to disentangle the spatial frequencies coming from the object or the attenuation of the OTF. 

\subsection{The mostly-unsupervised marginal estimation}

\begin{figure}
\centering
	\includegraphics[width=\columnwidth]{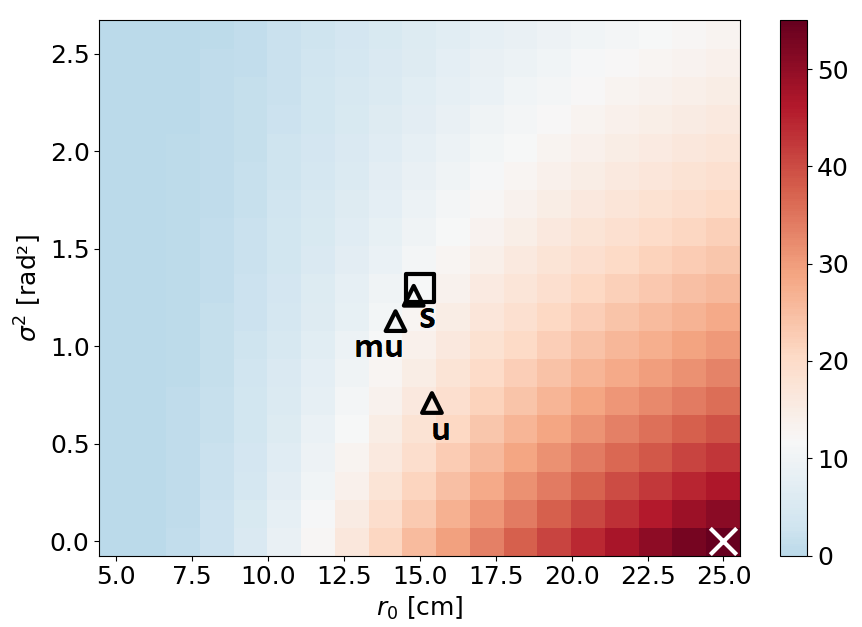}
    \caption{Strehl ratio corresponding to the different values of $r_0$ and $\sigma^2$ parameters. The true Strehl ratio (central square) is $\overline{S_R}=11.5\%$. Cross: minimum of $J_\text{JMAP}$ ($S_R=54.3\%$). Triangles: minima of $J_\text{marg}$ for the three different supervision levels (\emph{s}:supervised $S_R=11.4\%$, \emph{mu}:mostly unsupervised $S_R=11.0\%$, \emph{u}:fully unsupervised $S_R=17.1\%$).}
    \label{fig:strehl}
\end{figure}

\begin{figure}
\centering
	\includegraphics[width=\columnwidth]{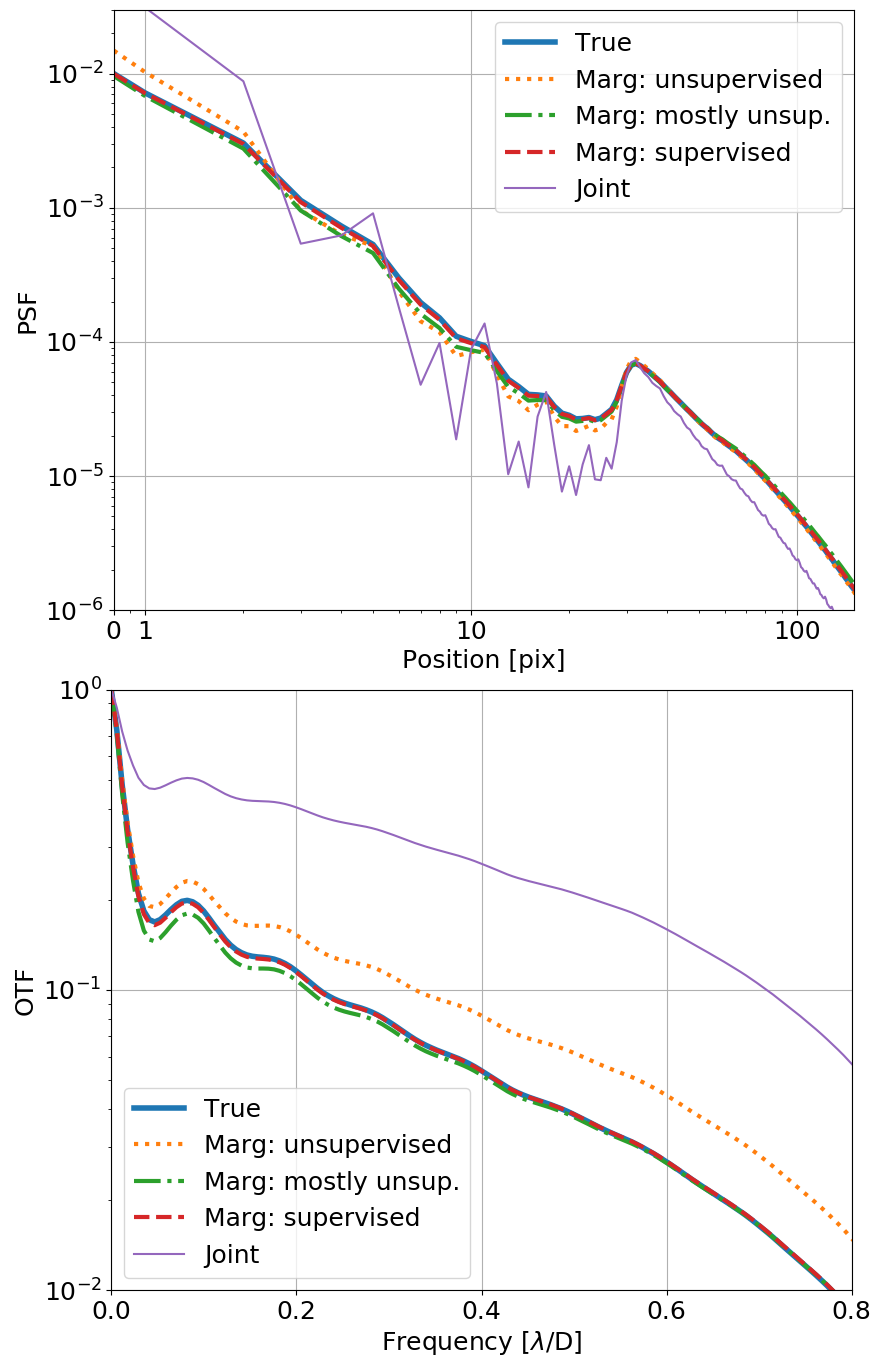}
    \caption{Top: True PSF (thick plain blue line), the marginal estimates for different supervision levels (dotted orange, dashed red, dashed-dotted green) and the joint estimate (thin violet plain). Bottom: corresponding OTFs. The frequency is normalised at $1$ at the telescope cutoff.}
    \label{fig:psf_otf}
\end{figure}

We have found empirically that the fully unsupervised marginal estimation leads to an over smoothing of the object by an over estimation of $p$. An operational method that should improve the result with respect to the fully unsupervised estimator is to supervise only the $p$ parameter, and to estimate simultaneously $\mu$ and $\rho_0$ along with the PSF parameters. Setting $p$ is to give to the algorithm the knowledge on the category of object that has been observed. We run a mostly-unsupervised marginal estimation where $p$ is fixed to the empirical guess value for asteroids $p=3.0$ and $(\mu,\rho_0)$ are estimated along with the PSF parameters $\bm{\gamma}$. The marginal algorithm solution is
\begin{equation}
    \left\{
    \begin{array}{rll}
     r_0 & = 14.2 & \text{ cm} \\ 
     \sigma^2 & = 1.13 & \text{ rad}^2 \\
     \mu & = 2.9\times 10^{-9} & \\
     \rho_0 & = 1.39 & \text{ pix}^{-1}
    \end{array}
    \right.
\end{equation}
The parameters $r_0$ and $\sigma^2$ are estimated with an error of $0.8$ cm and $0.17$ rad$^2$ respectively. These errors lead to a small under-estimation of the Strehl ratio (Fig.~\ref{fig:strehl}). The shape of the estimated PSF (Fig.~\ref{fig:psf_otf}) is of good quality, with a slightly higher halo than the true PSF, and consequently a slightly lower central peak. The OTF is also correctly estimated, with a small shift downwards at low frequencies.\\

\begin{figure}
\centering
	\includegraphics[width=.95\columnwidth]{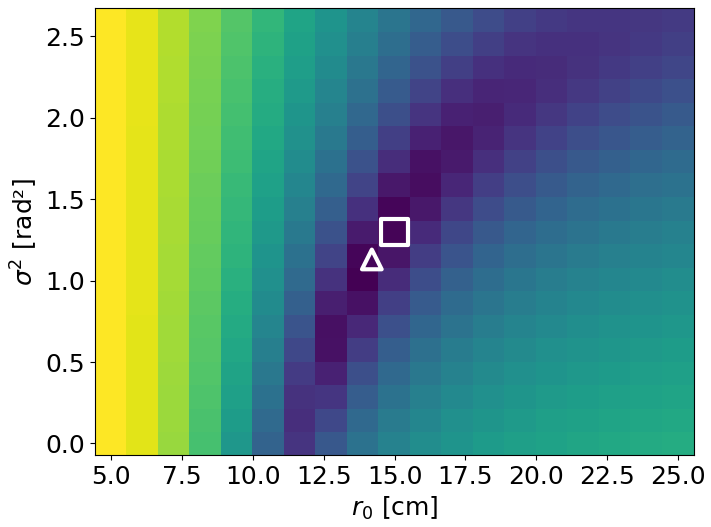}
    \caption{mostly-unsupervised marginal criterion $J_\text{marg}$ (log intensity scale) as function of the $r_0$ and $\sigma^2$ PSF parameters. The parameters $(\mu,\rho_0)$ are the ones estimated from marginal minimization, while object PSD power is fixed to the value $p=3$. Central square indicates the true PSF parameters, triangle indicates the minimum of the criterion.}
    \label{fig:Jmarg_most_u}
\end{figure}

\begin{figure}
\centering
	\includegraphics[width=\columnwidth]{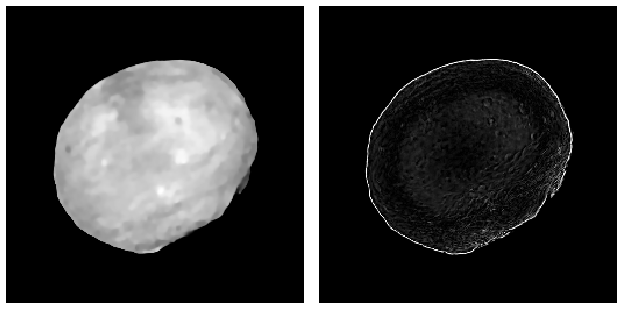}
    \caption{Left: deconvolution of the image using the PSF estimated by the mostly-unsupervised marginal estimation. Right: two times the absolute difference with the true object.}
    \label{fig:deconv_marginal_p_fixed}
\end{figure}

The overall shape of the PSF is suitable for deconvolution. Indeed, Fig.~\ref{fig:deconv_marginal_p_fixed} shows the deconvolution using the PSF estimated with mostly-unsupervised marginal approach. Similarly to the previous marginal cases, sharp edges, albedo and craterization details are restored. The deconvolution error is mainly located at the edges. The energy is more concentrated inside the object with respect to the border. The deconvolution error is $\varepsilon_\text{obj, marg mu} = 7.0\times 10^{-4}$. Both the PSF and the restored object are better estimated than in the unsupervised case.

\subsection{Results on SPHERE/Zimpol data}
\label{sec:sphere}

\begin{figure}
\centering
	\includegraphics[width=\columnwidth]{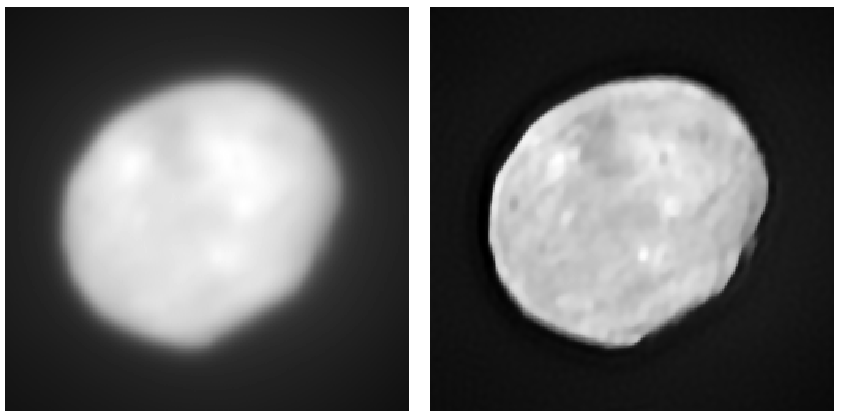}
    \caption{Left: VLT/SPHERE/Zimpol image of the asteroid Vesta on June 8$^\text{th}$ 2018. Right: Classical (non myopic) deconvolution using the PSF estimated by marginalisation.}
    \label{fig:sphere}
\end{figure}

The asteroid Vesta was observed as part of an European Southern Observatory (ESO) Large Program (ID 199.C-0074, PI P. Vernazza) targeting several Main Belt asteroids to study their albedo, shape, excavation volumes and surface craterization \citep{vernazza2020}. This Large Program provides unprecedented data to astronomers to study the evolution of Solar System objects. However the asteroid Vesta was previously visited by the NASA/Dawn interplanetary mission and has been resolved at high resolution \citep{russell2012,jaumann2012vesta} with the on-board framing camera \citep{sierks2011} down to $\sim 20$ meter per pixel. Vesta is consequently the perfect benchmark to test high resolution post processing techniques using ground-based telescopes images. The ESO Large Program is based on observations with the Zurich IMaging POLarimeter (ZIMPOL) instrument \citep{schmid2018sphere} and the N\_R filter of central wavelength $\lambda=645.9$ nm. This instrument is mounted on the Spectro Polarimetric High-contrast Exoplanet REsearcher (SPHERE) \citep{beuzit2019sphere} of the Very Large Telescope (VLT) observatory. It is equipped with the SAXO \citep{fusco2014} efficient adaptive optics system. In \cite{fetick2019vesta} we previously deconvolved the SPHERE/Zimpol images of the asteroid Vesta using a simple \cite{moffat1969} PSF model in the focal plane
\begin{equation}
    h(\vec{u}) = \frac{\beta-1}{\pi\alpha^2}\left( 1 + u^2/\alpha^2 \right)^{-\beta}
\end{equation}
The PSF parameters $(\alpha,\beta)$ were determined manually (supervised) for deconvolution and allowed astronomers to identify the main albedo features and craters from the images. We focus on the SPHERE observation of the night June 8$^\text{th}$ 2018 (Fig.~\ref{fig:sphere}, left). At this epoch, the asteroid orientation and illumination correspond to the OASIS simulated object of Fig.~\ref{sec:setup} (top left). In \cite{fetick2019vesta} the SPHERE image has been deconvolved with a Moffat PSF model of parameters $\alpha=4$ pixels and $\beta=1.5$. It provided accurate details with a resolution of $\sim 20 - 30$ km. The major issue comes from the difficulty of a Moffat model to describe the uncorrected halo of the AO PSF. The deconvolved objects consequently suffers from a residual halo.\\

\begin{figure}
\centering
	\includegraphics[width=\columnwidth]{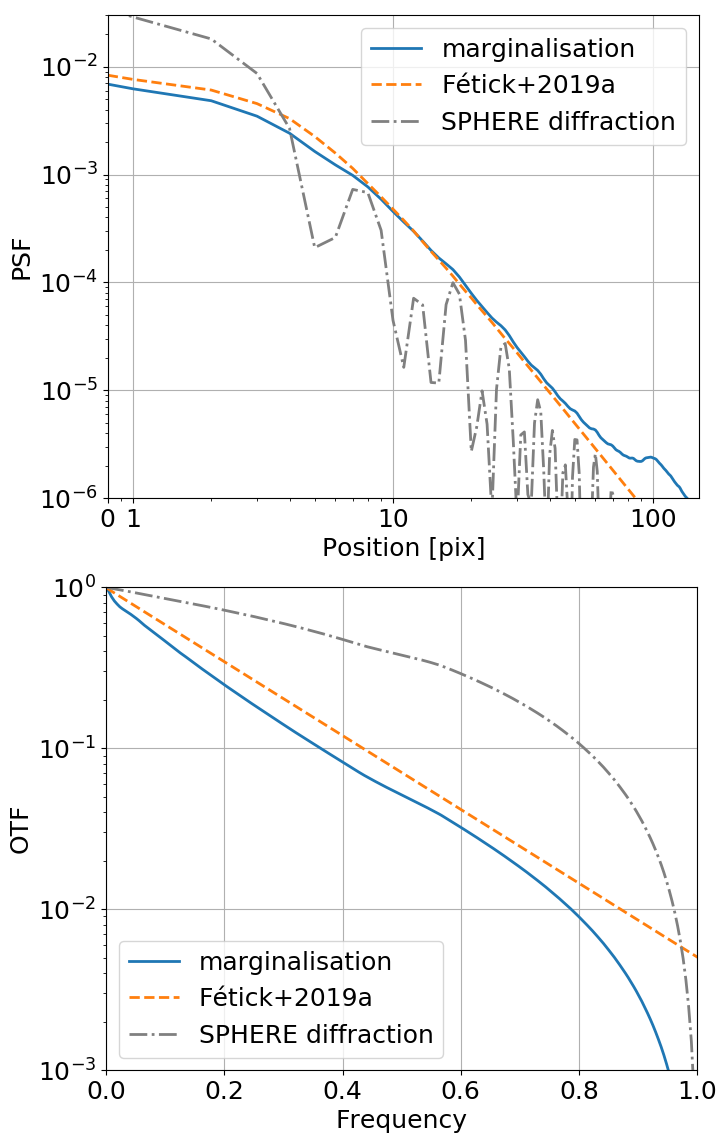}
    \caption{Top figure: PSF estimated for the SPHERE observation of Vesta on June 8$^\text{th}$ 2018. Plain line: using PSF model of Sect.~\ref{sec:setup} and mostly-unsupervised marginalisation. Dashed: Moffat model from \citet{fetick2019vesta}. Dashed-dotted: SPHERE diffraction limit. Bottom figure: corresponding OTFs.}
    \label{fig:sphere_psf}
\end{figure}

\begin{figure}
\centering
	\includegraphics[width=0.95\columnwidth]{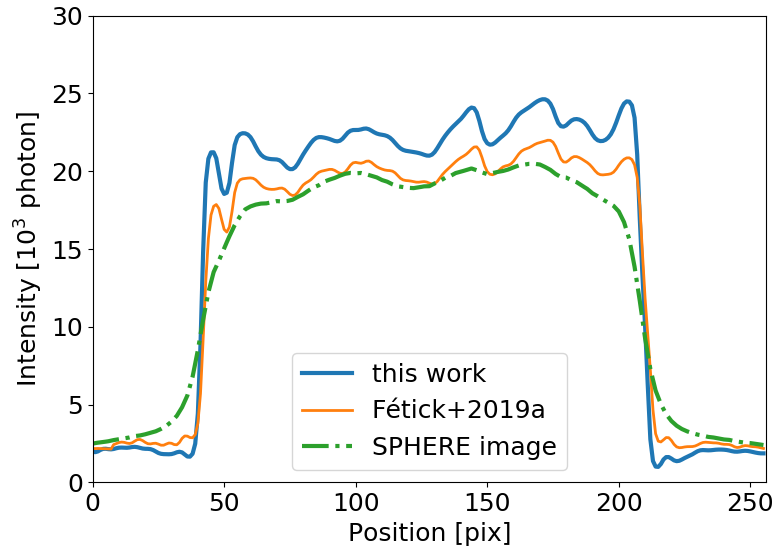}
    \caption{Slice of the asteroid Vesta for the SPHERE image (dashed-dotted), the deconvolution with a Moffat from \citet{fetick2019vesta} (thin plain) and the deconvolution with the marginally estimated PSF (thick plain).}
    \label{fig:sphere_slice}
\end{figure}

Since then, the two major advancements in post processing technique are the more accurate PSF model of Sect.~\ref{sec:setup} and the marginalisation to replace the manual -- supervised -- estimation of the PSF parameters. The observational fixed parameters given to our PSF model are summarised in Table \ref{tab:sphere}. To allow for high dynamics on the $(r_0,\sigma^2)$ parameters the marginal algorithm ranges are extended to $r_0\in [5,35]$ cm and $\sigma^2\in [0,4]$ rad$^2$. The algorithm is run in mostly-unsupervised mode, with object PSD power fixed to the value $p=3.0$. The solution at minimum is
\begin{equation}
    \left\{
    \begin{array}{rll}
     r_0 & = 32.1 & \text{ cm} \\ 
     \sigma^2 & = 2.78 & \text{ rad}^2 \\
     \mu & = 1.75\times 10^{-11} & \\
     \rho_0 & = 0.98 & \text{ pix}^{-1}
    \end{array}
    \right.
\end{equation}
The $\sigma^2$ value is consistent, however the $r_0$ estimation seems high. Indeed SPHERE AO system has a wide corrected area, the contribution of the turbulent part is of lower importance than the extended AO corrected core. The $r_0$ estimation could be improved using atmospheric telemetry as prior information for the marginal criterion. The corresponding PSF is shown on Fig.~\ref{fig:sphere_psf}. The global shape is quite similar to the one of the previous Moffat model. However the AO adapted PSF model shows stronger halo, which cannot be taken into account by the Moffat model. The Strehl ratio ($S_R=20\%$) is thus lower than the Moffat one ($S_R=29\%$). The slopes of the OTFs are similar, but the AO corrected model is globally shifted downwards starting from the low frequencies, due to the halo, since the Moffat model is not able to describe it correctly. Moreover the Moffat model does not take into account the telescope cutoff. The differences come from the limit of the Moffat model to describe an AO PSF with respect to the AO dedicated model. The marginal method adds consistency and robustness since it reduces the level of supervision in the PSF estimation. The deconvolution performed with the marginally estimated PSF is shown on Fig.~\ref{fig:sphere} (right). Details are restored making visible the main craters and albedo features. A slice of the asteroid is shown on Fig.~\ref{fig:sphere_slice}. The deconvolution with a Moffat model is able to retrieve the main details of the object, but an important fraction of the energy is contained in the residual object halo. The marginal estimation with the AO dedicated PSF restores similar details as the Moffat, but reduces the halo, the energy is more condensed inside the object and slightly sharper edges appear. For deconvolution with the Moffat PSF, $45\%$ of the energy remained outside the bounds of the OASIS -- true -- object, with the marginally estimated PSF the energy outside reduces to $38\%$.

\begin{table}
	\centering
	\caption{Parameters of the SPHERE/Zimpol instrument and fixed parameters of the PSF model.}
	\label{tab:sphere}
	\begin{tabular}{lcccr}
		\hline
		Parameter & Symbol & Value & Unit\\
		\hline\hline
		Primary diameter & D & 8 & m\\
		Secondary diameter & D' & 1.12 & m\\
		N\_R filter wavelength & $\lambda$ & 645.9 & nm\\
		Sampling & $s$ & 4.76 & N/A\\
		AO cutoff frequency & $f_\text{AO}$ & 2.5 & m$^{-1}$\\
		\hline
		Moffat elongation & $\alpha$ & 0.05 & m$^{-1}$\\
		Moffat power law & $\beta$ & 1.6 & N/A\\
		AO area constant & $C$ & $4\times 10^{-3}$ & rad$^2$m$^2$\\
		\hline
	\end{tabular}
\end{table}

\section{Conclusion and perspectives}
\label{sec:conclusion}

Post processing of astronomical images requires the knowledge of the PSF. For observational cases with no bright star in the field of view as PSF, a parametric model with few parameters is useful to perform post-processing such as deconvolution. However we  have  shown  that,  even  with the use of the recently developed dedicated AO PSF model, the joint estimator remains degenerated and always  converges  towards  the  sharpest  PSF  and  an over-smoothed deconvolved object.\\ 

In this paper, we have combined our parametric PSF model for AO-corrected images with the marginal PSF identification approach of \citet{Blanc2003b} and \citet{Blanco2011}. In supervised mode, it provides a very accurate estimation of the PSF and high quality deconvolution, similar to the ultimate deconvolution quality performed with the true PSF. The fully unsupervised mode exhibits a mild degeneracy with a trend to under-deconvolve. However, this marginal unsupervised estimation is still far more effective than the joint estimation. For practical applications, the marginal supervised mode can be relaxed to a mostly unsupervised mode. The remaining supervision targets only the power $p$ of the object's PSD. It means that the user provides to the marginal algorithm the information of which category of object has been observed: star field, artificial satellite or asteroid for example. We have also shown a good deconvolution quality for true VLT/SPHERE/Zimpol observations with adaptive optics.\\

The overall method we offer actually corresponds to a blind deconvolution, but separated in two consecutive steps: first the PSF estimation by the marginal method, and then the classical deconvolution with the marginally estimated PSF. It thus allows to perform blind deconvolution for different operational cases. Regarding the ESO Large Program on asteroids cited in this paper, approximately one third of the time was dedicated to PSF observations and often leading to poor deconvolution results. Thanks to the marginal method we are now able to perform better deconvolutions and to save an important fraction of the observing time. Even though we focused in this paper on deconvolution, the PSF estimated by the marginal approach can also be used in other post-processing techniques.\\

The marginal criterion used in this work assumes a Gaussian prior probability for the object, which allows negative values of the object. A perspective of this work is to impose the object's positivity to further improve the PSF identification.\\

Additionally, even though we focused on long-exposure imaging using a dedicated long-exposure AO corrected PSF model, this method could also be applied to short exposure imaging. To this end, the PSF model should be changed, for instance by parameterizing the PSF through the pupil phase, itself being described by a limited number of Zernike coefficients. \citet{Blanco2011} having already shown the possibility to estimate the defocus of an image with high precision, we are confident that this extension to short-exposure imaging would be successful.

\section*{Data availability}

The data underlying this article will be shared on reasonable request to the corresponding author.\\

\noindent The PSF model dedicated to AO observations is available at \emph{https://gitlab.lam.fr/lam-grd-public/maoppy}

\section*{Acknowledgements}

This work was supported by the French Direction G{\'e}n{\'e}rale de l'Armement (DGA) and Aix-Marseille Universit{\'e} (AMU). This work was supported by the Action Sp{\'e}cifique Haute R{\'e}solution Angulaire (ASHRA) of CNRS/INSU co-funded by CNES. This project has received funding from the European Union's Horizon 2020 research and innovation program under grant agreement No. 730890. This material reflects only the authors views and the Commission is not liable for any use that may be made of the information contained therein. This study has been partly funded by the French Aerospace Lab (ONERA) in the frame of the VASCO Research Project. This work has been partially supported by the ANR-APPLY program (ANR-19-CE31-0011). The authors are thankful to Pierre Vernazza for providing data related to his ESO Large Program ID 199.C-0074.


\bibliographystyle{mnras}
\bibliography{biblio}





\bsp	
\label{lastpage}
\end{document}